\documentclass[article,twocolumn,amsmath,floats,showpacs,nofootinbib]{revtex4}
\usepackage{graphicx}
\usepackage{textcomp}
\usepackage{hyperref}
\hypersetup{colorlinks=true}

\begin{document}
\title{Relaxation of nonequilibrium quasiparticles in a superconductor normal metal point contact}
\author{I. K. Yanson, N. L. Bobrov, L. F. Rybal'chenko, and V. V. Fisun}
\affiliation{B.I.~Verkin Institute for Low Temperature Physics and
Engineering, of the National Academy of Sciences
of Ukraine, prospekt Lenina, 47, Kharkov 61103, Ukraine
Email address: bobrov@ilt.kharkov.ua}
\published {(\href{http://fntr.ilt.kharkov.ua/fnt/pdf/13/13-11/f13-1123r.pdf}{Fiz. Nizk. Temp.}, \textbf{13}, 1123 (1987)); (Sov. J. Low Temp. Phys., \textbf{13}, 635 (1987)}
\date{\today}

\begin{abstract}The point-contact spectra of tantalum in the superconducting state, with $Ta$, $Cu$, and $Au$ counterelectrodes, have been studied. We discovered some new distinctive features, whose position on the $eV$ axis is determined by the critical power required for the injection of nonequilibriumquasiparticles. At this level of power the band gap $\Delta $ decreases abruptly in the vicinity of the contact. A correction to the point-contact spectrum, with the sign opposite to that of the usual correction, arises in the region of phonon energies. The maxima in the $Ta$ spectrum become sharper and their position on the energy axis becomes stabilized near the values $e{{V}_{ph}}=7.0$, 11.3, 15.5, and 18 $meV$, which correspond to low phonon group velocities $\partial \omega /\partial q\simeq 0$ in $Ta$. This is confirmed by the existence of corresponding flattenings on the dispersion relations $\omega (q)$ of lattice vibrations. Slow phonons are created near the $N-S$ interface in quasiparticle recombination and relaxation processes and cause a decrease in $\Delta $ and an increase in the differential resistance in the vicinity of $e{{V}_{ph}}$. An excess quasiparticle charge is accumulated in the region of the contact, producing a correction to the resistance, which decreases as $eV$, $T$, and $H$ increase. These mechanisms are particularly effective in dirty contacts, thus permitting phonon spectroscopy in the superconducting state even when the current flow occurs in a nearly thermal mode.

\end{abstract}

\maketitle

\section{INTRODUCTION}
The nature of the nonlinearity of the current-voltage characteristics (I-V curves) of point contacts between normal metals has been well studied.
In particular, the second derivative of the I-V curve has been shown to be proportional to the electron- phonon interaction (EPI) function at low temperatures. The study of the ${{{d}^{2}}V}/{d{{I}^{2}}}\;$ characteristics (point-contact spectra) for different metals and alloys is the domain of point-contact spectroscopy \cite{Yanson1, Yanson2}.

It is also known that if one electrode or both are in the superconducting ($S$) state, then the I-V curve of the point contact is essentially nonlinear for two reasons: 1) in the range of biases $eV\sim \Delta $ the nonlinearity is due to the band gap $\Delta $ in the spectrum of quasiparticle excitations \cite{Artemenko};  2) in the range of biases $eV\gg \Delta $ Joule heating can destroy superconductivity in the region of the point contact \cite{Iwanvshyn}. In three-dimensional point contacts, however, whose size \emph{d} is substantially smaller than the energy relaxation lengths for electrons and phonons, ${{\Lambda }_{e}}$ and ${{l}_{ph}}$, local equilibrium does not exist between the electrons and the lattice and the distribution functions of the quasiparticles are far from equilibrium functions. Under these conditions the quasiparticle distribution has a sharp edge at energies equal to the applied bias $eV$ and one could effect new effects associated with the spectral dependence of the density of electron and phonon states. Indeed, the first experiments performed in this field \cite{Khotkevich, Yanson3} showed that the fine structure of the nonlinearities of the I-V curves of $S-c-S$ contacts are associated with the characteristic energies of phonons, although the shape and exact position of the anomalies on the energy axis are often determined to a considerable degree by the shape of the contact and the purity and structural perfection of the metal in the region of the contact. Distinct and reproducible phonon features were observed in the spectra of $S-c-N$ contacts  \cite{Yanson4}. The advantages of $S-c-N$ contacts consist in the absence of the Josephson effect, which results in additional nonlinearities of the I-V characteristics, as well as in less lattice heating in the region of the contact in the case of a normal metal with a high thermal conductivity. The disadvantages are a certain complication of the interpretation of the spectra, because of the possible manifestation of the electron-phonon interaction of the normal metal, as well as the necessity to take into account the differences in ${{p}_{F}}$ and ${{v}_{F}}$ in the contacting metals when making quantitative estimates of the shape and intensity of the spectra and the diameter of the contact, \cite{Shekhter} because under real conditions an $S-c-N$ contact always is a heterocontact.

Theory \cite{Khlus1, Khlus2} and experiment \cite{Yanson4} have shown that for pure $S-c-N$ contacts the point-contact spectrum basically is similar to the spectrum in the normal ($N$) state and differs from the latter only by an additional contribution due to inelastic processes with the emission of phonons under Andreev reflection of electrons, forming an excess current.
For $\Delta \ll eV,\ \ \delta \left( eV \right)$  [where $\delta \left( eV \right)$ is the width of the spectral lines of the point-contact EPI function] the contribution of these processes is small.

On the other hand, in a previous study \cite{Yanson3} we found that in dirty $S-c-S$ and $S-c-N$ contacts the shape of the phonon features in the point-contact spectra differs substantially from that in normal metals. Moreover, a distinct phonon structure in the $S$ state was observed even when it was absent in the $N$ state.

In this work we consider the idea that relaxation of quasiparticle excitations in a superconductor can determine the phonon structure of point-contact spectra to a considerable measure, especially in the case of contacts with a sufficiently high concentration of elastic scatterers. Clearly, the deviation of the electron and phonon subsystems from equilibrium in the region of the contact cannot be ignored. When current is passed through an $N-S$ contact an imbalance arises between the populations of the electron and hole branches of the quasiparticle spectrum in the superconducting electrode \cite{Lemberger}. One result of this is a reverse current of quasiparticles, which leads to an increase in the resistance in the range of low biases \cite{Peshkin}. Beside this, when the current in the semiconductor increases the total concentration of quasiparticle excitations increases in the vicinity of the contact and suppresses $\Delta $, and when a critical concentration is reached, brings about a transition to a spatially inhomogeneous nonequilibriumstate \cite{Galitskii}. In this paper we present experimental facts, which in our view confirm the substantial role of nonequilibrium phonons with low group velocities in the formation of new features of the point-contact spectra during the transition to the S state. The preliminary results were given earlier in a brief communication \cite{Yanson5}.
\section{FABRICATION AND CHARACTERISTICS OF CONTACTS}
\begin{figure}[]
\includegraphics[width=8cm,angle=0]{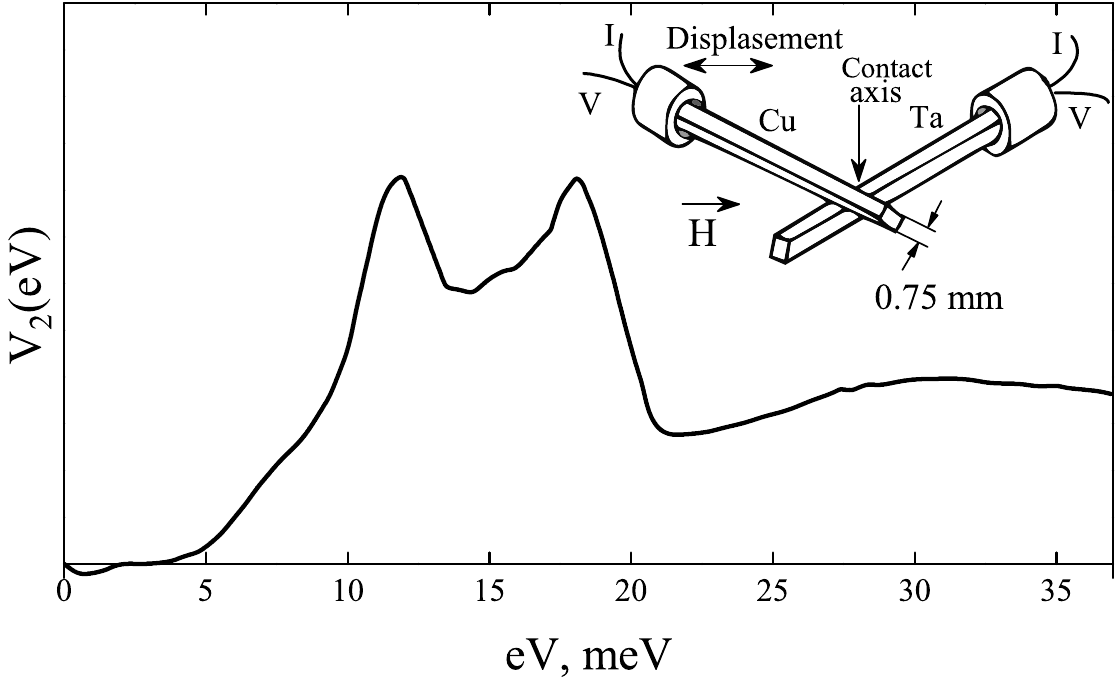}
\caption[]{Point-contact spectrum of a $Ta-Cu$ heterocontact in the N state (the effective value of the modulated voltage at zero bias was ${{V}_{1.0}}=406\ \mu V$, $T=1.72\ K$, ${{R}_{0}}=73\ \Omega $, $d=47\ \text{{ }\!\!{\AA}\!\!\text{ }}$; at a bias of 20 mV the differential resistance increases by 7\%). The inset illustrates the geometry of the experiment.}
\label{Fig1}
\end{figure}

For the superconducting electrode we chose tantalum, which can be used in conjunction with metals such as copper or gold to form point contacts that have a resistance in the range from a few ohms to several hundred ohms and possess spectra with a relatively low background and distinct phonon maxima in the $N$ state. One such spectrum, serving as a "certificate" for each sample, is given in Fig. \ref{Fig1}, which also shows the experimental geometry in the inset.

\begin{figure}[]
\includegraphics[width=8cm]{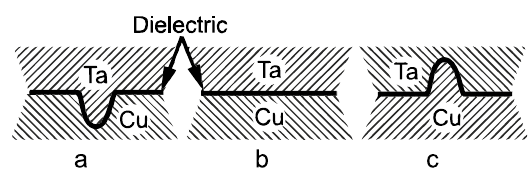}
\caption{Some possible configurations of Ta-Cu heterocontacts.}\label{2}
\label{Fig2}
\end{figure}

A detailed study of the point-contact spectra of such samples in the N state was carried out in Ref.\cite{Bobrov1}. Two different methods were employed to fabricate the point contacts. In the first method, electrodes were made to touch in liquid helium, thus creating an electrical contact with a resistance of the order of 1-10 k$\Omega $, which was then decreased smoothly to the desired value during the passage of a shaping current at a bias of a few tenths of a volt. The metal in the region of the contact heats up and softens at such high biases, so that the area of the contact spot increases under the applied force and the region of the contact is partially purified of oxides and structural faults. In the second method electrical shaping at high biases was not carried out and the region of the contact was purified as a result of microdisplacements of the electrodes relative to each other in the plane of contact (shift method, see Ref. \cite{Yanson1}). On the whole, the observed characteristics depended little on the method of contact fabrication, but a high-quality sample is much easier to obtain by means of electrical shaping.

\begin{figure}[]
\includegraphics[width=8cm,angle=0]{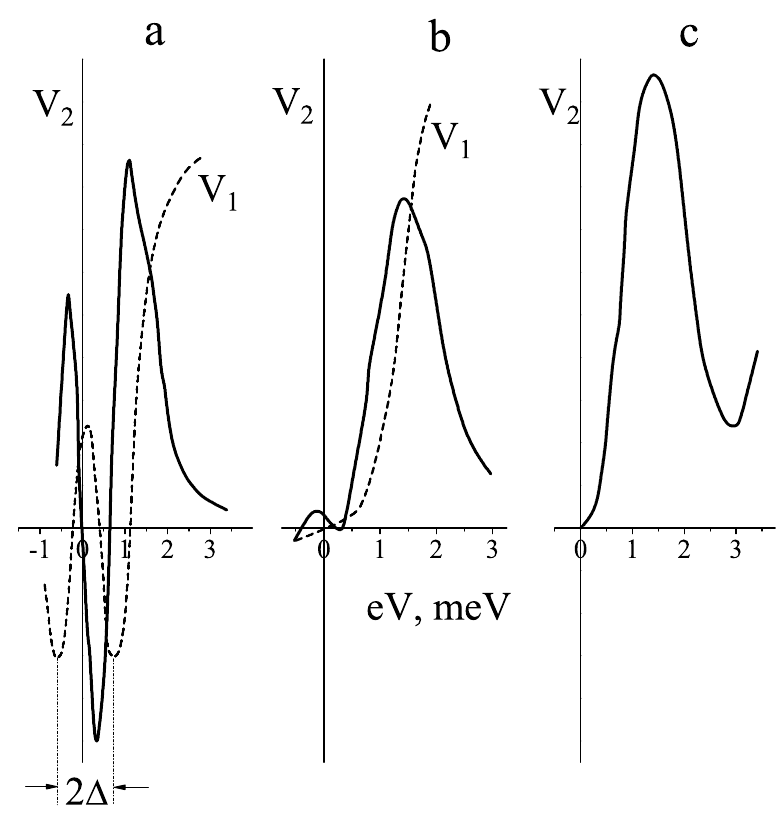}
\caption[]{Initial portions of the point-contact spectra of Ta-Cu:
a) ${{R}_{0}}=24.5\ \Omega ,\quad T=1.65\ K$;
\\b) ${{R}_{0}}=26.5\ \Omega ,\quad T=1.5\ K$;
\\c) ${{R}_{0}}=33\ \Omega ,\quad T=1.65\ K$.}
\label{Fig3}
\end{figure}

One can conceive of different variants of configurations and structure of real contacts (Fig. \ref{Fig2}).
Unfortunately, the structure of a contact cannot be determined by direct methods (e.g., observation under an electron microscope, nor can a contact be fabricated with a prescribed geometry and structure. This is prevented by the extremely small size $\left( d\approx 20-200{\AA} \right)$ and the fact that the geometry must in principle be three-dimensional. Diverse electrical measurements thus are the sole source of information. Utilizing them, we can learn a good deal about the structure of the contact without destroying it. In the point-contact spectra of the contacts shown in Fig. \ref{Fig2}(a, c), for example, the predominant features are the peaks that correspond to the phonon spectrum of the metal, which predominantly fills the region of constriction. We note that for $Ta$, paired with a noble metal, spectra with an appreciable contribution from the latter are extremely rare. Unfortunately, structures of the type shown in Fig. \ref{Fig2} (a, b) cannot be distinguished for the given pair. The causes of this were analyzed in detail in Ref.\cite{Bobrov1}.
The I-V characteristic and its derivatives in the range of biases of the order of $\Delta$ can be used for an analysis of the structure of an $S-c-N$ contact.
In particular, a contact with a structure similar to the model in Fig. \ref{Fig2}(b) has tunneling-like I-V curves that display a minimum of ${dV}/{dI}\;$ at $eV=\Delta $ (see Fig. \ref{Fig3}a, where the voltage ${{V}_{1}}$ is proportional to ${dV}/{dI}\;$ and ${{V}_{2}}\sim {{{d}^{2}}V}/{d{{I}^{2}}}\;$).

The presence of defects and impurities in the boundary layer leads to additional electron reflection (in comparison with that due to the difference in the electron characteristics), which, on the one hand, changes the shape of the I-V curve, making it more similar to that of ideal tunnel contacts, and on the other hand sharply decreases the excess current ${{I}_{exc}}$ that is observed as the difference between the I-V characteristics in the S and N states in the range $eV>\Delta $ (Ref.\cite{Blonder}). We selected contacts whose ${{I}_{exc}}$  were more than half the theoretically expected value \cite{Zaitsev1}\footnote{For Ta-Cu contacts the theory \cite{Zaitsev1} gives ${{I}_{exc}}(T=0)=(1.1-1.2)/\Delta R$ for the values b=0.87 and ${{{v}_{F}}Ta}/{{{v}_{F}}Cu=0.65-0.87}\;$. The excess current is roughly the same for most of the contacts whose spectra are given in this paper.}  and, therefore, the interface was sufficiently clean in the case of the structure in Fig. \ref{Fig2}b. An additional check that a large number of elastic scatterers was not present in the region of the contact was made on the basis of the intensity of the point-contact spectra in the N-state, which should decrease appreciably for ${{l}_{i}}\ll d$ $({{l}_{i}}$ is the impurity mean free path) (Ref.\cite{Yanson2}). The electron elastic mean free path in our contacts lies in the interval\footnote{The diameter of the $Ta-Ta$ contacts was evaluated from the formula $d=37.6/{{R}^{{1}/{2}\;}}\ (nm)$ and that of the Ta-Cu contacts, from $d=39.74/{{R}^{{1}/{2}\;}}\ (nm)$, where $R$ is the resistance in ohms (we assumed that $\ {{{p}_{F}}Ta}/{{{p}_{F}}Cu}\;=0.87$ and ${{{v}_{F}}Ta}/{{{v}_{F}}Cu=0.65}\;$; cf. Ref. \cite{Bobrov1}); the value obtained was $d\lesssim 60\ {\AA}$.}  $d\lesssim {{l}_{i}}\lesssim {{l}_{i,o}}$, where ${{l}_{i,o}}$ is the mean free path in bulk Ta $(\approx900 {\AA})$. We note that the use of a purer material in itself does not guarantee that cleaner contacts are obtained. Radical changes in  the fabrication technology are required to obtain cleaner contacts. In Fig. \ref{Fig3} (b,c) the initial portions of the point-contact spectra correspond to point contacts that have a geometry close to that of the model in Fig. \ref{Fig2}a.

The main measurements were made on $Ta-Cu$ contacts. The characteristics of the $Ta-Au$ contacts were similar. Besides the I-V curve and its derivatives, which were recorded by the standard modulation method, we also recorded the deviation $\delta I(V)$ of the I-V curve from Ohm's law by means of a bridge circuit. When the bridge is balanced the $\delta I(V)$ curve is horizontal in the region of small $eV$ since the I-V curve of the point contact is ohmic in the N state. This is not always convenient since as $eV$ increases the $\delta I(V)$ characteristic abruptly passes into the negative region (when the gain is sufficiently high). In a number of cases, therefore, the bridge was slightly unbalanced so that the straight line corresponding to Ohm's law was at a certain angle to the abscissa axis. Regardless of the recording method used, the excess current is the difference $\delta {{I}_{S}}-\delta {{I}_{N}}$ and the calibration of the ordinate axis is preserved.
\section{TEMPERATURE DEPENDENCES OF POINT-CONTACT SPECTRA}
\begin{figure}[]
\includegraphics[width=8cm,angle=0]{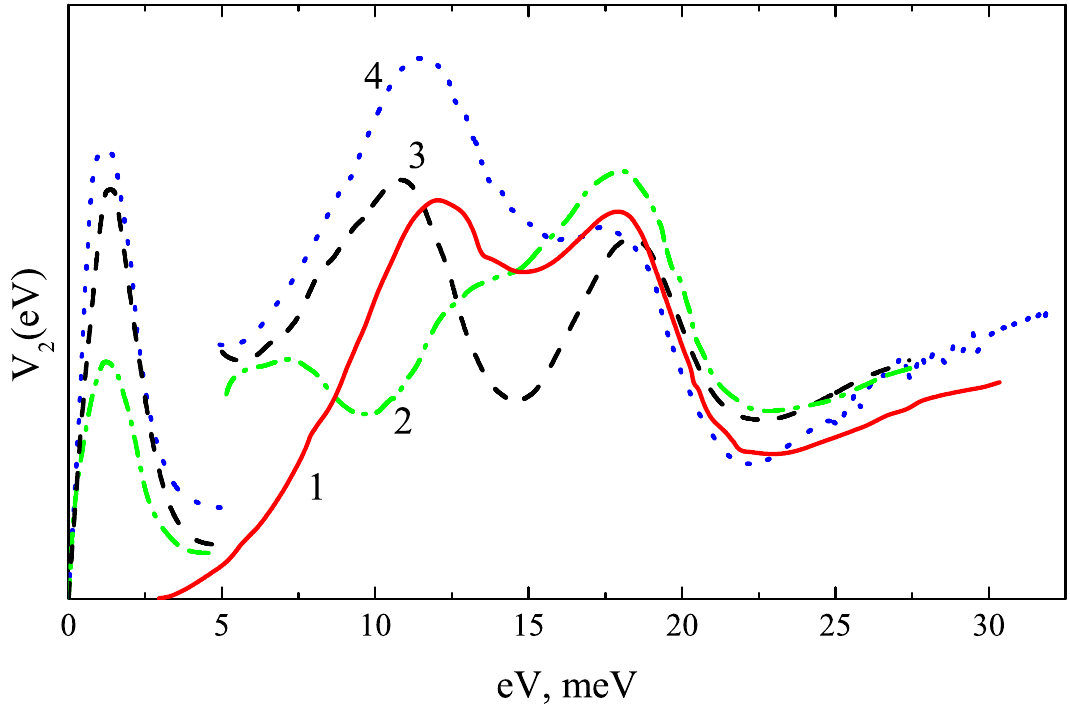}
\caption[]{Point-contact spectra of Ta-Cu near ${{T}_{c}}$=4.26(1),  4.22 (2), 4.20 (3), 4.13 K (4) (${{R}_{0}}$=70 $\Omega $, H=0).}
\label{Fig4}
\end{figure}

\begin{figure}[]
\includegraphics[width=8cm,angle=0]{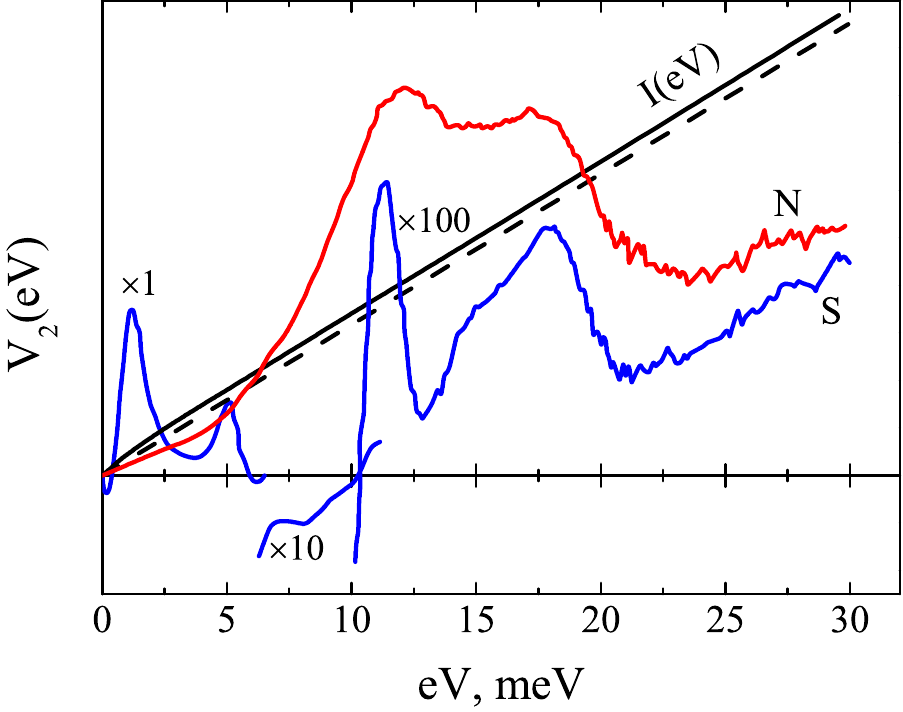}
\caption[]{Current-voltage characteristic and point-contact spectra of a Ta-Cu contact in the N and S states [${{R}_{0}}$=18 $\Omega $, $\delta R/R_0(10\to 30\ mV)$=2.7\%, $T$=5 and 2.83 K for the N and S states, respectively, ${{V}_{1}}$ (10$mV)=$628 $\mu V$]. The ${{V}_{2}}$ scale is changed by a factor of 10 (-10 dB) and 100 (-20 dB) by changing ${{V}_{1}}$ by a factor of $\sqrt{10}$ and 10, respectively.}
\label{Fig5}
\end{figure}

Signs of superconductivity in the point-contact spectra of $S-c-N$ contacts usually appear at a temperature ${{{T}'}_{c}}$ that is lower than the ${{T}_{c}}$ of bulk $Ta$ (${{T}_{c}}=4.48\ \text{K}$). From Fig. \ref{Fig4}
we see that beside a rapid increase in the intensity of the band-gap anomaly in the region of small values of $eV$, the point-contact spectrum undergoes considerable changes in a narrow range of temperatures near ${{T}_{c}}$. These changes consist in a decrease in the intensity at the beginning of the TA peak ($eV\simeq 11\div 12\ meV$), and in some cases so does that of the LA peak $eV\simeq 18\ meV$. Such behavior cannot be a result of the destruction of superconductivity in the region of the contact and hence, the spectrum of the N state should be reproduced. As the temperature is lowered further in the case of relatively clean contacts the intensity of the peaks is restored approximately and the change in the point-contact spectrum is smoother, causing the phonon maxima to undergo systematic peaking and the spectrum move downward relative to the spectrum in the normal state, to negative values of ${{d}^{2}}V/d{{I}^{2}}$ (Fig. \ref{Fig5}).
The EPI feature at $eV$=7 $meV$, which corresponds to the low-frequency mode in $Ta$ (Ref.\cite{Woods}) and usually appears as a weak shoulder in spectra of the N state \cite{Bobrov1}, in this case becomes a maximum. Similar peaking is also undergone by the feature at $eV\simeq 15 meV$, which is noticeable only in the spectra of high-quality contacts in the N state \cite{Bobrov1} and is absent from the N spectrum (Fig. \ref{Fig5}).

\begin{figure}[]
\includegraphics[width=8cm,angle=0]{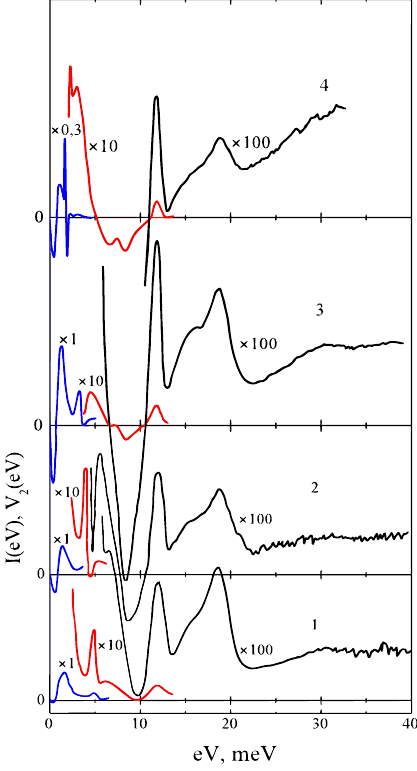}
\caption[]{Effect of variation of the resistance on the point-contact spectrum of a Ta-Cu heterocontact in the S state:\\
1) ${{R}_{0}}=87\ \,\Omega ,\quad T=1.44\,\ K$;\\
2) ${{R}_{0}}=41\,\ \Omega $;\\
3) ${{R}_{0}}=28.5\ \,\Omega $;\\
4) ${{R}_{0}}=13.4\,\ \Omega ,\quad {{T}_{2-4}}=1.63\,\ K.$\\
The multipliers next to the curves indicate the change in the scale of the ordinate axis. The excess currents for curves 1-4 was 1.21, 1.11, 1.21, and 0.93$\Delta /{{R}_{0}}$, respectively $(\Delta =0.7\,meV)$.}
\label{Fig6}
\end{figure}

While in the N states the position of the main $\text{TA}$ and LA peaks in the EPI point-contact spectrum varies within the limits 11.5-12.5 and 17-18 $meV$, respectively, for various contacts, in the S state the position of these peaks were recorded at biases of 11.3 and 18 $meV$ to within an error of the order of $\pm0.1 meV$. The effect of stabilization of the position of peaks on the $eV$ axis during the transition to the S state is especially noticeable for contacts whose size is comparable to the mean free path (i.e., for low-resistance and/or dirty contacts). The effect of the diameter of the contact on the shape of point-contact spectra in the S state is shown in Fig. \ref{Fig6},
which gives the second derivatives of the I-V characteristic of the same contact whose resistance was successively decreased by electrical shaping. It is seen that the lower the resistance, the more the point-contact spectrum moves lower along the ordinate axis, entering the negative region, and the more pronounced the peaking of the maximum.

\section{DISCONTINUOUS VARIATION OF THE PROPERTIES OF A SEMICONDUCTOR NEAR A CONTACT}
\begin{figure*}[t]
\includegraphics[width=16.7cm,angle=0]{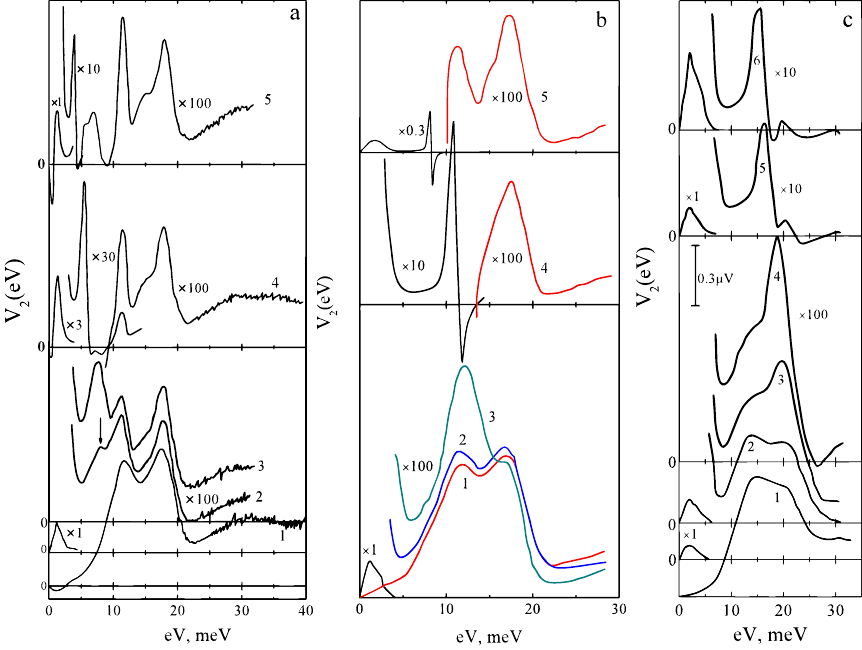}
\caption[]{\\a) Formation of a nonequilibrium peculiarity (indicated by an arrow above curve 2) near the low-frequency mode of the phonon spectrum ($eV=7\ meV$) for a $Ta-Cu$ contact, ${{R}_{0}}=64\ \Omega ,\ \, d=50\text{{ }\!\!{\AA}\!\!\text{ }}$, $T$=4.8(1), 3.9(2), 3.0 (3), 2.25 (4), and 1.4 K(5).\\
b) Formation of a nonequilibrium peculiarity near the TA peak of the phonon spectrum ($eV=11.5\ meV$) for a $Ta-Cu$ heterocontact, ${{R}_{0}}=202\ \ \Omega ,\ \,d=28\text{{ }\!\!{\AA}\!\!\text{ }}$, $T$=4.8(1), 4.15 (2), 3.5 (3), 2.6 (4), and 1.6 K(5).\\
c) Formation of a nonequilibrium peculiarity near the LA peak of the phonon spectrum ($eV=17.5\ meV$) for à $Ta-Ta$ heterocontact, ${{R}_{0}}=270\ \ \Omega ,\ \,\ d=23\text{{ }\!\!{\AA}\!\!\text{ }}$, $T$=4.8(1), 3.94 (2), 3.63 (3), 3.34 (4), 2.8 (5), and 2.55 K(6).}
\label{Fig7}
\end{figure*}

The point-contact spectra in the S state also have peculiarities (at $eV=5\ meV$ in Fig. \ref{Fig5} and at $eV=2.5-5\ meV$ in Fig. \ref{Fig6}), whose position on the V axis depends on the resistance of the contact, the temperature, or the external magnetic field.
These peculiarities are absent from the spectra in the N state. They have a tendency to appear near characteristic phonon energies when the temperature is lowered below ${{T}_{c}}$. Figure \ref{Fig7}a,b,c
shows cases when such peculiarities appear near the low-frequency mode and the maxima of the density of states of the $\text{TA}$ and $\text{LA}$ phonons, respectively.

\begin{figure}[]
\includegraphics[width=8cm,angle=0]{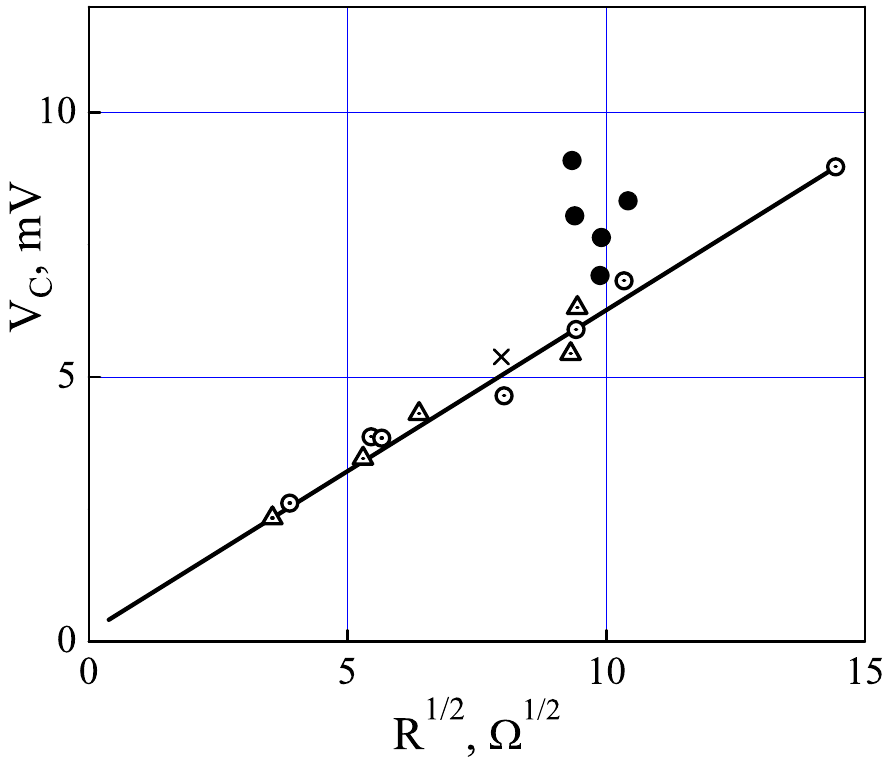}
\caption[]{Critical temperature as a function of the resistance of the contact at $T$=2 K: $(\Delta )$ data for a $Ta-Cu$ contact, obtained in one cycle of measurements; $(\bigodot)$ data corresponding to those given in Fig. \ref{Fig9}. The scatter of the points $(\bullet)$ near $R\approx 100\ \ \Omega $ is due to contacts with tunnel barriers.}
\label{Fig8}
\end{figure}

\begin{figure}[]
\includegraphics[width=8.5cm,angle=0]{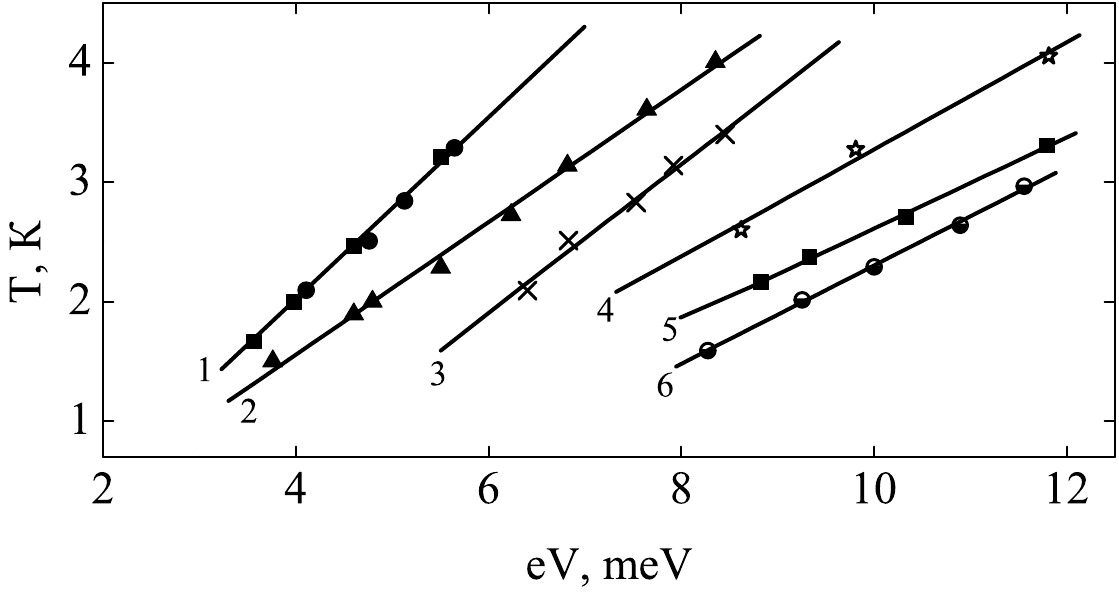}
\caption[]{Critical voltage as a function of the temperature for $Ta-Cu$ heterocontacts, R=32(1), 64(2), 90(3), 107(4), 95(5) and 210\ $\Omega $(6).}
\label{Fig9}
\end{figure}

\begin{figure*}[]
\includegraphics[width=17cm,angle=0]{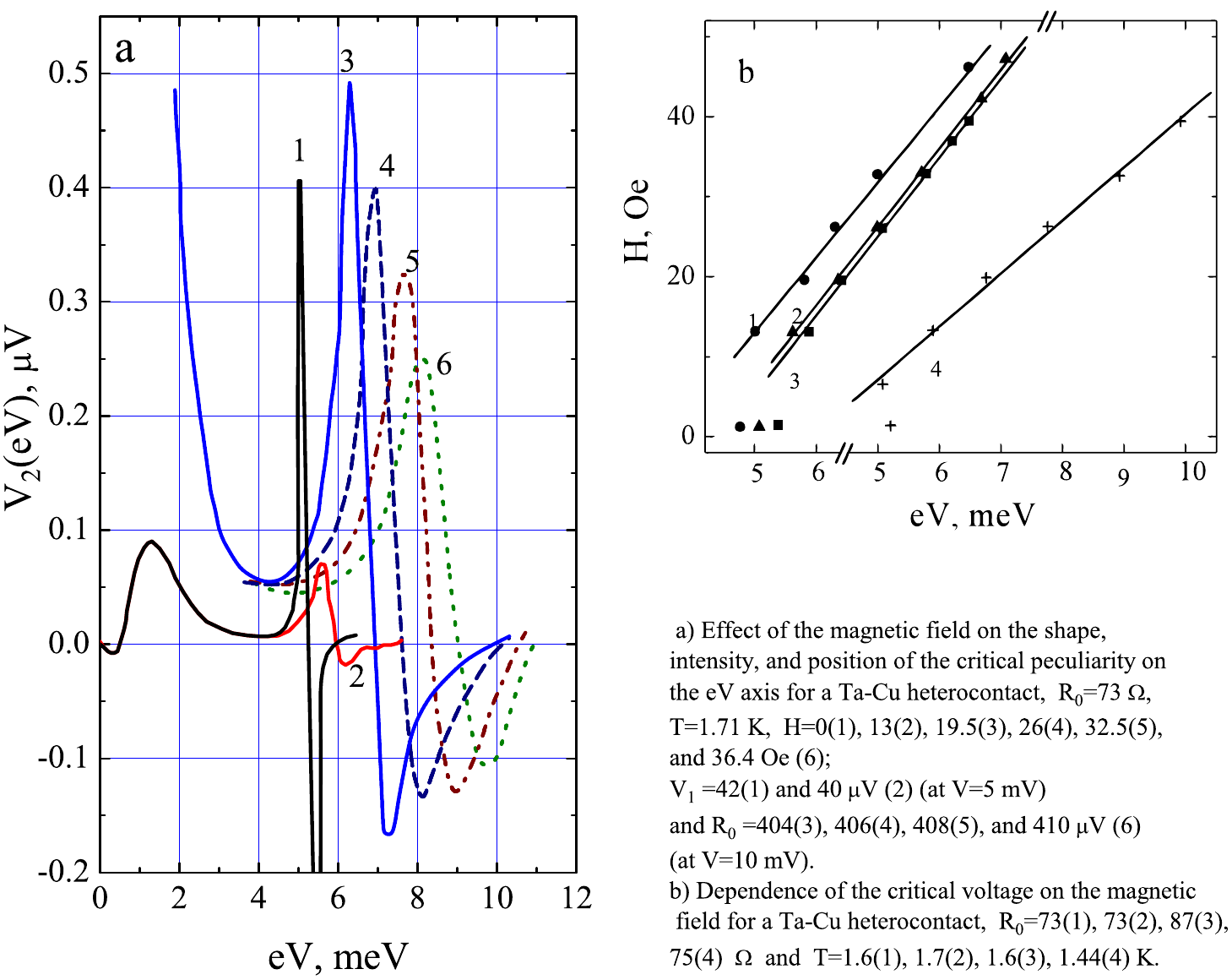}
\caption[]{}
\label{Fig10}
\end{figure*}

The higher the resistance of the contact, the higher the biases at which a peculiarity arises.
At a fixed temperature the voltage at which a peculiarity is observed is proportional to ${{R}^{1/2}}$ (Fig. \ref{Fig8}),
 which corresponds to a constant critical power ${{{P}_{c}}=V_{c}^{2}}/{R\simeq\text{const}}\;$ ($\simeq 0.4\mu \text{W}$ for $Ta$ at 2 K). The critical power \underbar{increases} with increasing temperature or external magnetic field, whose orientation is not of major importance (Figs. \ref{Fig9}, \ref{Fig10}).

This rules out the interpretation of a peculiarity as a result of the destruction of the superconductivity by the temperature or the field and unambiguously points to the essentially nonequilibrium origin of the peculiarity. Indeed, if the peculiarity is due to an abrupt transition of a nonequilibrium to a new state when a certain critical quasiparticle concentration is reached, the steady-state density of the quasiparticles depends on their relaxation rate, which increases with the temperature and field. The injection current ${{I}_{c}}={{V}_{c}}/R$ should increase so as to compensate for this increase.

\begin{figure}[]
\includegraphics[width=8cm,angle=0]{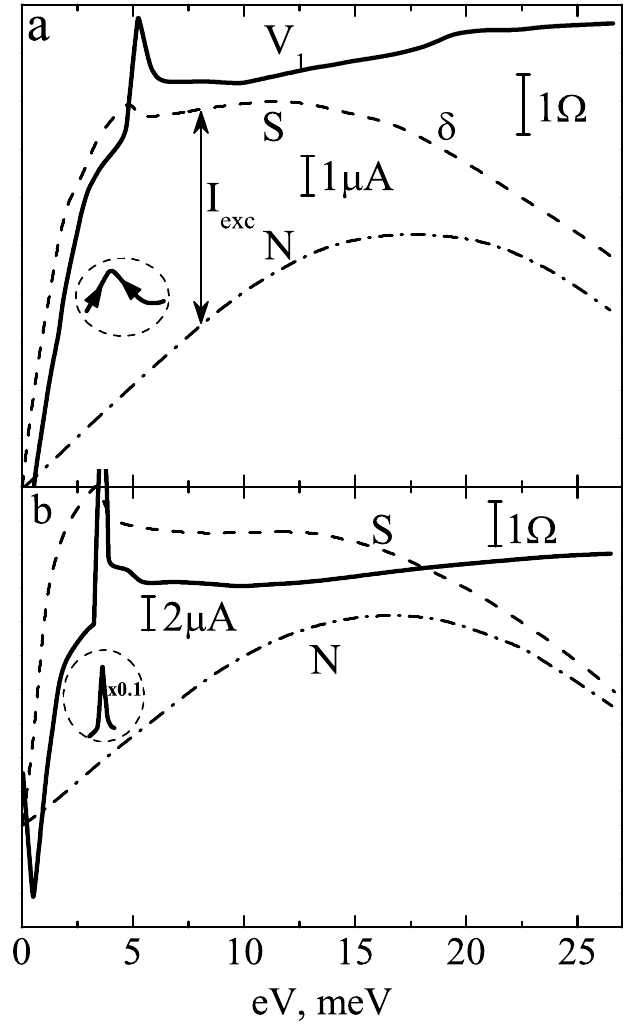}
\caption[]{Excess current and differential resistance as functions of $eV$ for two $Ta-Cu$ heterocontacts, ${{V}_{1}}\propto {dV}/{dI}\;$ (N is the deviation of the I-V characteristic from Ohm's law in the normal state and $S$ is the deviation in the superconducting state). The initial portions of the $N$ curves correspond to Ohm's law:
a) ${{R}_{0}}=62.5\ \Omega ,$ $T$=1.48 K;\\
b) ${{R}_{0}}=24.5\ \Omega ,$ $T$=1.65 K.
The inset to Fig. \ref{Fig11}a shows the reversibility of the I-V characteristic in the region of the critical peculiarity for a contact with ${{R}_{0}}=26.1\  \Omega ,$ T=1.5\ K.}
\label{Fig11}
\end{figure}

The shape of the peculiarity corresponds to an abrupt decrease in the excess current and increase in the differential resistance by a value of the order of several per cent (Fig. \ref{Fig11}).
 The I-V characteristic is not usually observed to have hysteresis in the region of a peculiarity (see inset to Fig. \ref{Fig11}). Since the excess current changes only slightly after a jump, the band gap in the region of the contact remains virtually equal to its previous value.

The critical density of excess quasiparticles, which cause a peculiarity to appear, can be evaluated if we take into account that the minimum volume of superconductor that can undergo a phase transition should not be smaller than the coherence length $\xi$. Then ${{n}_{c}}\sim {{J}_{c}}\tau $ where ${{J}_{c}}\sim 3{{P}_{c}}/4\pi \Delta {{\xi }^{3}}$ is the rate of quasiparticle generation per unit volume and $\tau$ is the relaxation time, which in the simplest case is equal to the time taken by the quasiparticles to escape from the region under consideration, i.e., $\tau \sim \xi /{{v}_{F}}$. We assumed that for typical values of ${{V}_{c}}$ the inelastic mean free path is greater than $\xi$ and the flight is ballistic. When estimating ${{J}_{c}}$ we considered that an energy $\sim \Delta$. In other words, there is a multiplication of quasiparticles by reabsorption of nonequilibrium phonons and electron collisions, so that eventually the majority of quasiparticles have excess energy $\sim \Delta$. Since $\xi =\xi (l)$, and decreases abruptly along with $l(eV)$ at biases near the maxima of the density of phonon states, the density of the nonequilibriumquasiparticles increases unevenly with the bias and decreases abruptly at the characteristic phonon energies. Moreover, phonons with low group velocities, which are reabsorbed effectively in the contact, correspond to these energies. This explains which the formation of a nonequilibrium peculiarity in point- contact spectra is usually "linked" to regions of more rapid decrease of $l(eV)$, i.e., to the maxima of the EPI functions (Fig. \ref{Fig7}).
\section{Ta-Ta CONTACTS}
\begin{figure}[]
\includegraphics[width=8cm,angle=0]{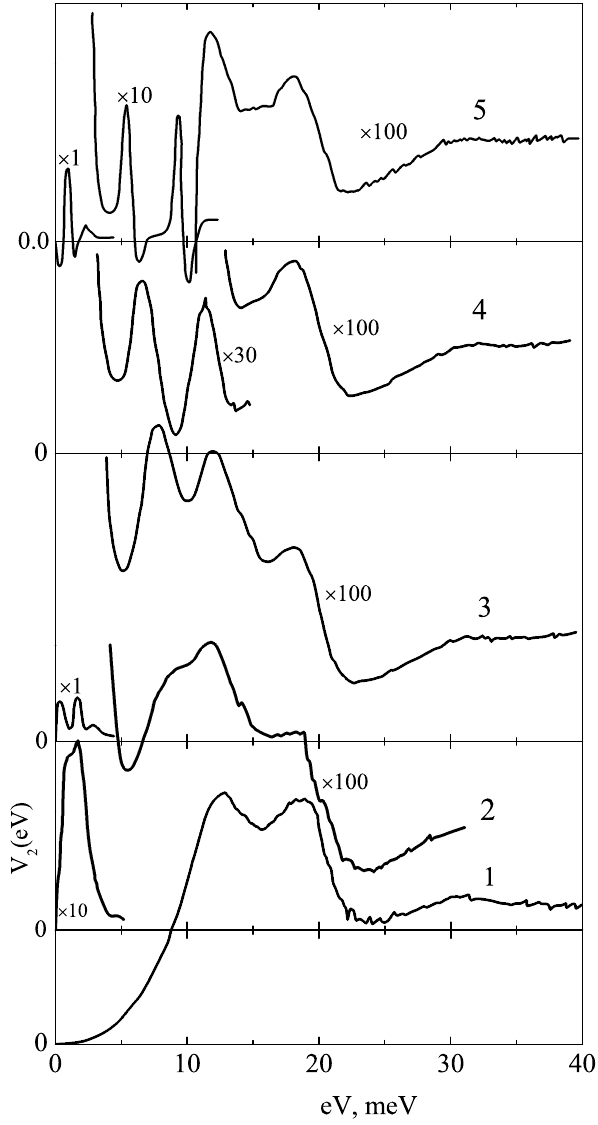}
\caption[]{Point-contact spectra of a $Ta-Ta$ homocontacts, ${{R}_{0}}=64\ \ \Omega ,$  $T$=4.6 K (curve 1, $N$ state), 3.5(2), 3.25(3), 2.74(4), 2.0 K(5); ${{V}_{1,0}}=0.446\ mV,$  ${{V}_{2,\max }}=0.62\ \mu V$ (for curve 1).}
\label{Fig12}
\end{figure}

\begin{figure}[]
\includegraphics[width=8cm,angle=0]{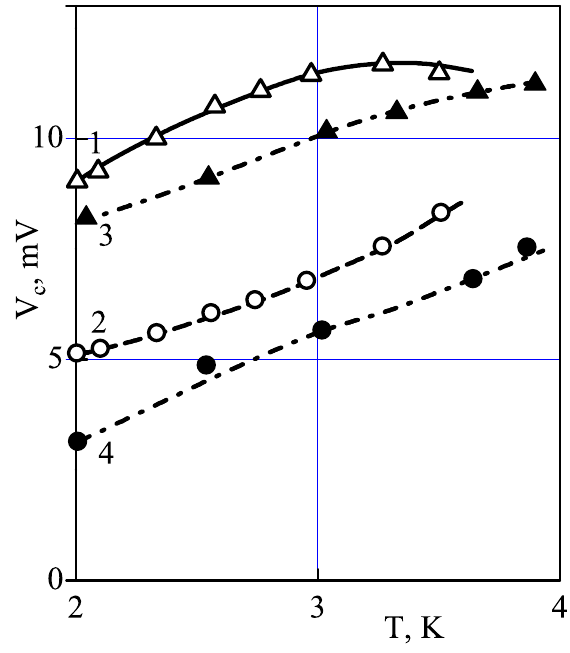}
\caption[]{Critical voltages as functions of the temperature for two nonequilibrium peculiarities in the point-contact spectra of two $Ta-Ta$ homocontacts:\\
(1,2): ${{R}_{0}}=64\ \Omega $, $\delta {R}/{{{R}_{0}}\,\left( 20\ mV \right)=7.5\%}\;$, ${{I}_{exc}}=1.2{\Delta }/{{{R}_{0}};}\;$\\
(3,4): ${{R}_{0}}=54\ \Omega $, $\delta {R}/{{{R}_{0}}\,\left( 20\ mV \right)=8.3\%}\;$, ${{I}_{exc}}=1.5{\Delta }/{{{R}_{0}}.}\;$ The critical power of the first peculiarity at $T$=2.0 K is $0.44\ \mu $W.}
\label{Fig13}
\end{figure}

We also studied the temperature evolution of point-contact spectra of $Ta$ homocontacts at temperatures below ${{T}_{c}}$.
Typical results, which are given in Fig. \ref{Fig12},
unambiguously indicate that the critical peculiarities are due to a change of state of the tantalum edge and not to the proximity effect near the interface with copper (or gold). This was to be expected since the diameter of the contact is substantially smaller than both ${{\xi }_{0}}$ and  $\xi_{N}=h{{v}_{F}}/2\pi T$ and the proximity effect is suppressed to a considerable degree. From Fig. \ref{Fig12} we see that two peculiarities, corresponding to each of the edges, arise in $S-c-S$ homocontacts. Their position on the $V$ axis as a function of the temperature is shown in Fig. \ref{Fig13}
 for two different contacts. As for $S-c-N$ contacts, the critical power, which is proportional to ${{V}_{c}}^{2}$, increases with rising temperature and its value at $T$=2 Ê for the first critical peculiarity coincides with the critical power in an $S-c-N$ contact. These results can be compared with the critical peculiarities observed in Ref.\cite{Khotkevich} for tin $S-c-S$ contacts, which have the same shape but a higher intensity. Since the values of the band gap in $Ta$ and $Sn$ are approximately equal, ${{P}_{c}}(Sn)/{{P}_{c}}(Ta)\approx {{\xi }^{3}}(Sn)/{{\xi }^{3}}(Ta)\approx 25,$ which is in agreement with the value ${{P}_{c}}=9\ \mu W$, which is observed for tin at the same relative temperature. Assuming that for $Ta$ at $T$=2 K we have $\Delta =0.67\ meV,\ {{\xi }_{0}}=900{\AA},$ and ${{v}_{F}}=0.74\cdot {{10}^{8}}cm/\sec ,$ we obtain the critical concentration ${{n}_{c}}\approx 0.16\cdot {{10}^{18}}c{{m}^{-3}}.$
This is the typical density of nonequilibrium quasiparticles at which an abrupt transition occurs in a spatially inhomogeneous state and during tunnel injection \cite{Dynes}.

\section{DISCUSSION OF EXPERIMENTAL RESULTS. EXCESS CHARGE OF QUASIPARTICLESIN REGION OF CONTACT}
The qualitative explanation for the observed variations in the point-contact spectra in the transition from one of the electrodes to the $S$ state is based on taking into account the fact that the electrons and phonons in the region of the contact deviate substantially from equilibrium. Quasiparticles with maximum energy $eV\approx \hbar {{\omega }_{D}}\gg \Delta $ are injected into the superconductor through the $N-S$ interface. These quasiparticles populate the electron-like or hole-like branch of the spectrum of excitations, depending on the polarity of the applied voltage. The excitations relax \cite{Kaplan} \footnote{The electron-phonon scattering time in a point-contact usually is several times the fundamental time ${{\tau }_{ep}}$ (Ref. \cite{Kaplan}) because of the difference between the electron distribution function and the equilibrium form of the function.} comparatively quickly, in a time of the order of ${{\tau }_{ep}}(\varepsilon )$, emitting phonons and accumulating in a layer of the order of $\Delta $ above the top of the band gap. Further relaxation of the residual imbalance between the occupation and the total concentration of excess quasiparticles occurs fairly slowly, in a time of the order of ${{\tau }_{0}}\sim {{\tau }_{ep}}(\Delta )$ (Ref. \cite{Kaplan}), during which the excitation manages to diffuse into the superconductor to a depth  $\lambda_{Q} \sim {{\left( {{l}_{i}}{{l}_{\Delta }} \right)}^{{1}/{2}\;}}$, where  ${{l}_{\Delta }}={{v}_{F}}{{\tau }_{0}}$. Since the potential difference decreases at a distance of the order of \emph{d} from the plane of the contact, the chemical potential of quasiparticles in the region of the contact is not equal to the chemical potential of pairs, in much the same way as in tunnel $S-I-N$ contacts. An excess quasiparticle charge of density Q*(r) arises in this case along with a reverse current (or additional voltage), which increases the resistance of the contact more than Q*(d) does. The factors that decrease the magnitude of the imbalance (inelastic scattering on phonons, superconducting current, or an external magnetic field when scattering occurs on impurities) reduce the reverse current and the resistance of the contact.

The changes that occur in the spectrum upon transition to the $S$ state, as shown in Figs. \ref{Fig4}-\ref{Fig6}, can be explained from this standpoint. Indeed, since the additional resistance decreases as a result of the increase in the frequency of electron-phonon scattering with increasing bias $eV$, the I-V characteristic (i.e., its second derivative) has the sign opposite to the sign due to ordinary inelastic processes, which also occurs in the $N$ state. The effect should decrease for all spectra as the bias $eV$ increases, as is indeed observed experimentally. Then, near ${{T}_{c}}$ the imbalance relaxation is slowed down by a factor of ${T}/{\Delta }\;$ and this possibly explains the strong effect observed in this temperature range. On the whole, the spectrum in the $S$ state should lie below the spectrum in the $N$ state, which is also consistent with experiment.

\begin{figure}[]
\includegraphics[width=8cm,angle=0]{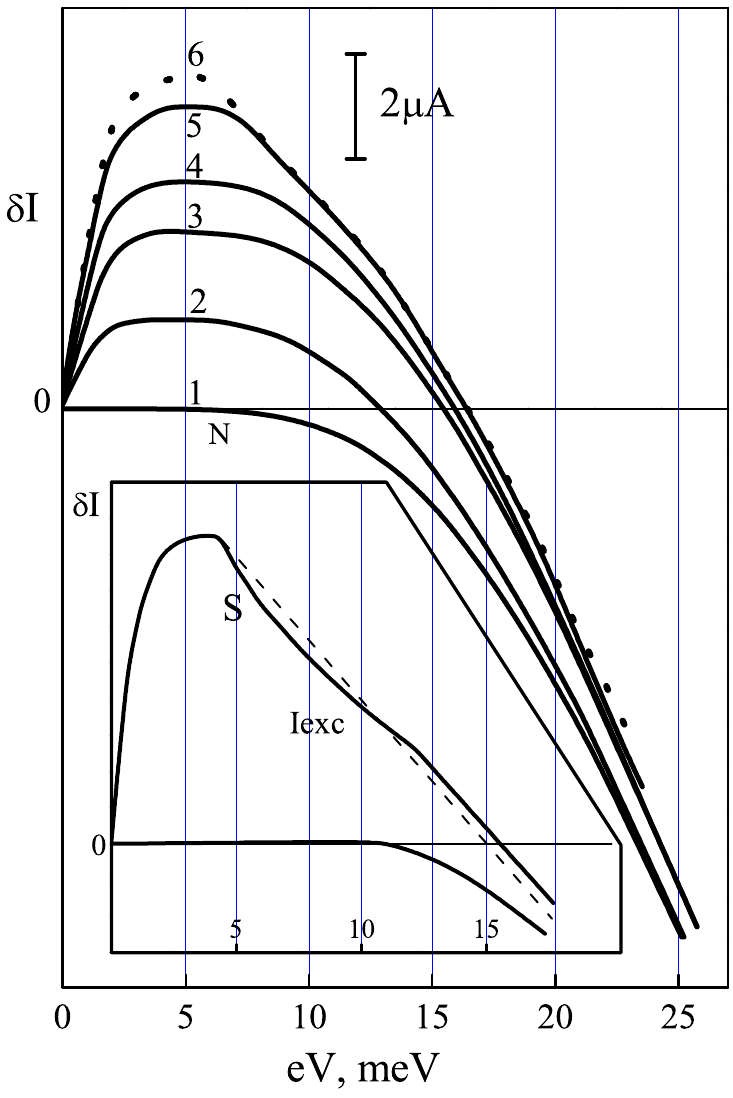}
\caption[]{Dependences $\delta I(eV)$ for a $Ta-Cu$ contact, ${{R}_{0}}=61\ \ \Omega,$ $T$=4.2 K (curve 1, \emph{N} state), 3.93(2), 3.81(3), 3.46(4), 2.63(5), and 2.03 K(6) (\emph{S} state). The inset shows $\delta I(V)$ for the heterocontact whose spectra are shown in Fig. \ref{Fig5}.}
\label{Fig14}
\end{figure}

The foregoing discussion shows that the excess current, which is defined as the difference between the I-V characteristics in the $S$ and $N$ states, should depend in an extraordinary way on the magnetic field (in the presence of elastic scatterers), the temperature, or the bias, which increase the rate of charge-imbalance relaxation. It is seen clearly from the inset to Fig. \ref{Fig14}
that the dependence of the excess current on $eV$ first has a positive curvature and only at high biases, at which the influence of the reverse current can be ignored, is the curvature negative in accordance with the theory \cite{Khlus1,Khlus2}. Figure \ref{Fig14} also shows a family of $\delta I(V)$ curves, taken at various temperatures for another contact. If the nonequilibrium effects were insignificant, then for any bias ${{I}_{exc}}=\delta {{I}_{S}}-\delta {{I}_{N}}$ and along with it $\delta {{I}_{S}}$ (since $\delta {{I}_{N}}$ is virtually temperature-independent) should increase with decreasing temperature because the band gap grows, since the excess current is proportional to the value of $\Delta $  in the direct proximity (at a distance of the order of \emph{d}) of the contact. We see that for the lowest temperature (curve 6) in a certain range of biases the $\delta I({{T}_{1}})$ curve lies below $\delta I({{T}_{2}})$ when ${{T}_{2}}>{{T}_{1}}$. From this we can conclude that, generally speaking, the excess current measured in point contacts is always smaller than the value predicted by the theory, which does not take nonequilibrium effects into account \cite{Artemenko,Zaitsev2}. In our case, however, these corrections are small and the experimental excess current is close to the theoretical value.

\begin{figure}[]
\includegraphics[width=8cm,angle=0]{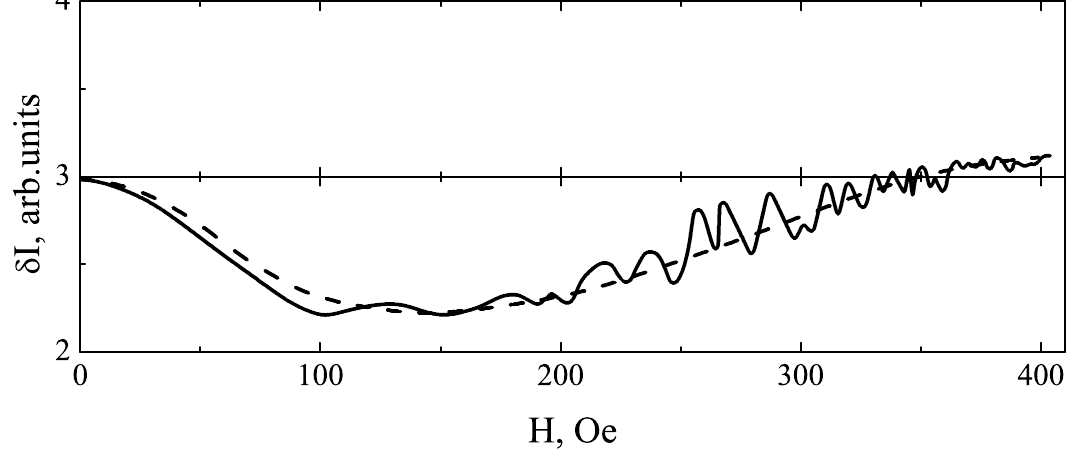}
\caption[]{Excess current at a bias $eV=11.5\ meV$as a function of the magnetic field for a $Ta-Cu$ heterocontact, ${{R}_{0}}=6\ \ \Omega ,$    $T=1.63\ K.$}
\label{Fig15}
\end{figure}

We find more evidence of the influence of nonequilibrium quasiparticles by examining the dependence of the excess current, at a fixed bias, on the external magnetic field (Fig. \ref{Fig15}).
If we leave the oscillations aside, we can note that in sufficiently high fields (of the order of several tenths of ${{H}_{cm}}$) $\delta I$ and ${{I}_{exc}}$ \underline{increase} with increasing field, often reaching values that exceed the zero-field values. In the equilibrium situation the magnetic field causes a decrease in $\Delta $, and together with it, the excess current.

We shall try to describe the form of the contribution to the point-contact spectrum from the energy relaxation of the quasiparticle charge imbalance. Suppose that the spreading of quasiparticles over long distances in comparison with \emph{d} is diffusive in nature: $\lambda_{Q}\gg {{l}_{i}}\gtrsim d$. The quasiparticle charge density then decreases with depth in the superconductor according to the law \cite{Peshkin}.
\begin{equation}
\label{eq__1}
Q^*(r)=Q^*(a)\frac{a}{r}{{e}^{{-r}/{{{\lambda }_{Q}}}\;}},
\end{equation}
Where ${{\lambda }_{Q}}={{\left( \frac{1}{3}\ {{l}_{i}}{{l}_{Q}} \right)}^{1/2\;}},\ \ {{l}_{Q}}={{v}_{F}}{{\tau }_{Q}},\ \ d=2a$.

Following Ref.\cite{Peshkin}, we obtain
\begin{equation}
\label{eq__2}
{{\tau }_{Q}}l=2\pi aQ^*(a){{\lambda }_{Q}}({{\lambda }_{Q}}+a)
\end{equation}
(\emph{I} is the total current through the contact),
\[{p_{Q}}={{{I}_{rev}}}/{I\simeq \frac{3}{2}\frac{a}{{{l}_{i}}}\left( 1-\frac{a}{{{\lambda }_{Q}}} \right),}\]
and the second derivative of the I-V characteristic of the point contact  has the form

\[\frac{1}{R}\frac{dR}{dV}=-\frac{3}{2}\frac{a}{{{l}_{i}}}\frac{{{d}^{2}}}{d{{V}^{2}}}\left( \frac{{{V}a}}{{{\lambda }_{Q}}} \right).\]

\begin{figure}[]
\includegraphics[width=8cm,angle=0]{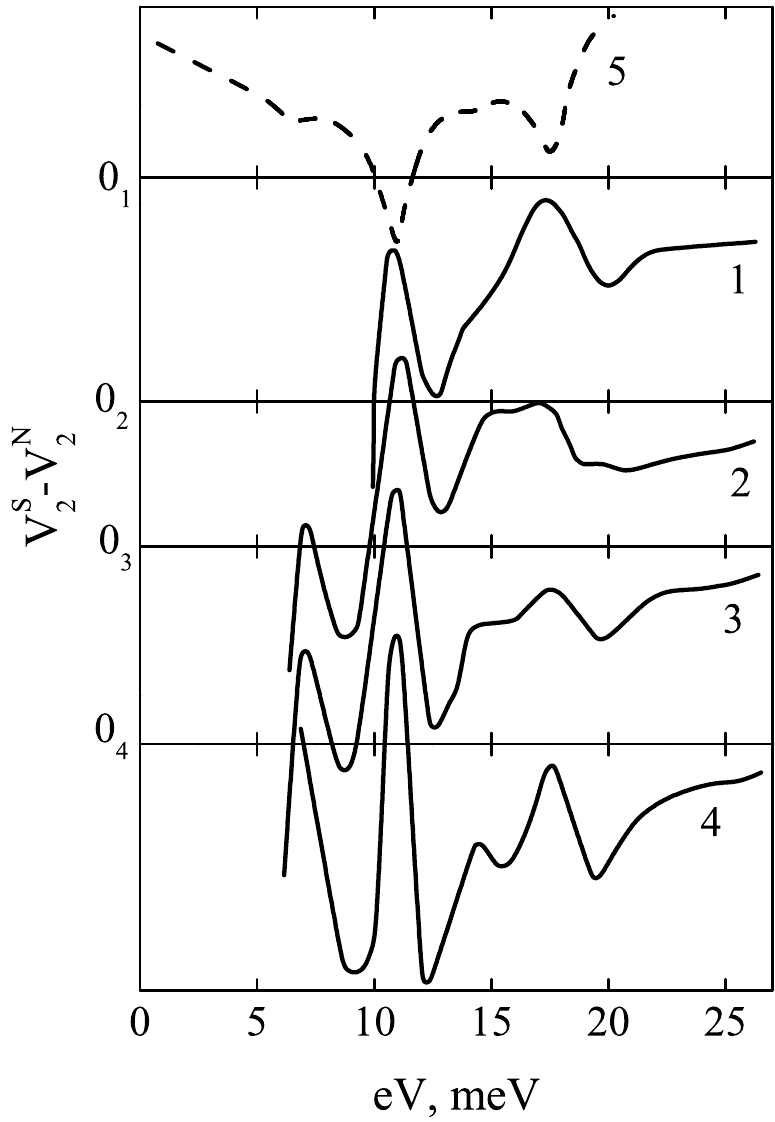}
\caption[]{Difference of the point-contact spectra in the $S$ and $N$ states for $Ta-Cu$ heterocontacts with ${{R}_{0}}$=202 (1), 64 (2), 73(3), and 85 $\Omega $ (4) and the dependence corresponding to formula (3) (curve 5).}
\label{Fig16}
\end{figure}

Here, as above, ${{\lambda }_{Q}}$ and ${{\tau }_{Q}}$ are the average values for quasiparticles in a layer that is a distance $eV$ above the Fermi energy. Ignoring the $K$- factor, we write
\[{{l}_{Q}}^{{-1}/{2}\;}=\frac{1}{eV}\int\limits_{0}^{eV}{\frac{d\varepsilon }{{{\left( {{v}_{F}}{{\tau }_{Q}} \right)}^{{1}/{2}\;}}},}\]
where ${{\tau }_{Q}}(\varepsilon )$ is the imbalance relaxation time for quasiparticles of energy $\varepsilon $, which was calculated in Ref.\cite{Kaplan}. Finally, for the second derivative of the I-V characteristic by $T=0$ we obtain
\begin{equation}
\label{eq__3}
\frac{1}{{{R}_{0}}}\frac{dR}{dV}\simeq -\frac{\sqrt{3}}{2}\frac{{{a}^{2}}}{{{l}_{i}}^{{3}/{2}\;}{{v}_{F}}^{{1}/{2}\;}}\frac{d}{dV}\left[ \frac{1}{{{\tau }_{Q}}^{{1}/{2}\;}\left( eV \right)} \right].
\end{equation}
Figure \ref{Fig16}
 shows the difference of the point-contact spectra in the $S$ state (at $T=1.5\ K$ and the $N$ state for several contacts (curves 1-4). For comparison we also give the graph of - ${d}/{dV\left[ {{\tau }_{Q}}^{{-1}/{2}\;}(eV) \right]}\;$, which was calculated from formula (21) of Ref.\cite{Kaplan}, using the tabulated function $g(\omega )$ (Ref.\cite{Rowell}). Clearly, the spectral peculiarities on the experimental curves $V_{2}^{S}-V_{2}^{N}$ differ qualitatively from the predictions of the model taking the imbalance into account. Indeed, at the characteristic phonon energies 7.0,  11.3, 15.5, and 18 $meV$ the experimental curves have maxima, which sometimes go even into the region of positive values, whereas the theory, which on the whole correctly predicts a negative sign for the superconducting contribution to the spectrum, gives minima. This discrepancy cannot be eliminated by refining the model and compels us to look for a different explanation of the observed peculiarities.

\section{REABSORPTION OF NONEQUILIBIRUM PHONONS}
Analyzing the ensemble of experimental facts, we come to the conclusion that the reabsorption of phonons by electrons is of decisive importance for the formation of specific peculiarities in the point-contact spectra of contacts with superconducting electrodes. In a previous paper \cite{Yanson5} we noted that the multiplication of quasiparticles through the reabsorption of nonequilibrium phonons is important only for an estimation of the total concentration of excitations, on which the band gap and the excess current depend. The phonons reabsorbed are primarily those that slowly leave the region of the contact, i.e., phonons with low group velocities $\partial \omega /\partial q$ and corresponding frequencies, at which the density of states is a maximum. Precisely the selection of phonons by the criterion $\partial \omega /\partial q=0$ determines the effect of stabilization of the positions of the maxima and their peaking during the transition to the $S$ state, while in the $N$ state phonons participate in inelastic one-phonon backscattering processes over a wider region of phase space and the position of the maxima in the spectrum can change according to the orientation of the axis of the contact relative to the crystallographic axes. When elastic scatterers are introduced into the region of the contact the point-contact spectra in the $N$ state as a rule become broader, e.g., as a result of smearing of the anisotropy. Reabsorption of nonequilibrium phonons in the superconductor, leading to an increase in the total number of quasiparticles and to a decrease in the band gap near the constriction, gives rise to a qualitatively new effect. Elastic scattering of phonons hinders their escape and accentuates the effect of selection. For dirty contacts and/or large contacts, therefore, the maxima on the energy axis of the point-contact spectra in the $S$ state are very narrow and their position on the energy axis is located exactly near singular points of the phonon spectrum: $\partial \omega /\partial q=0$. In accordance with this, the dispersion curves for phonons in $Ta$ should have flattenings at frequencies that correspond to the energies of the maxima $e{{V}_{ph}}$: 7.0, 11.3, 15.5, and 18.0 $meV$. The last three energies correspond to frequencies where $\partial \omega /\partial q=0$ for the principal crystallographic directions in $Ta$, according to data on the inelastic scattering of neutrons \cite{Woods}. Moreover, the curve of $T-$ phonons in the [111] direction in Ref.\cite{Woods}  is observed to have an inflection, which can correspond entirely to the maximum ${{V}_{2}}(V)$ at $e{{V}_{ph}}=7\ meV$ of lower intensity. More exact neutron experiments should support this prediction, which was made on the basis of an analysis of point-contact spectra.

In the \emph{N} state phonon reabsorption leads to a background in the point-contact spectra. The experimental data indicate that the greater the background, the more markedly the spectrum in the $S$ state differs from the spectrum in the $N$ state. This can also be considered as indirect evidence of the role of nonequilibrium phonons in the change in the shape of the point-contact spectrum in the transition to the $S$ state.
\section{CONCLUSIONS}
Let us summarize the studies on the mechanism by which the point-contact spectra of $Ta-Ta$, $Ta-Cu$, and $Ta-Au$ contacts are transformed during the transition of $Ta$ to the $S$ state. When $eV\gg \Delta$ the deviation from equilibrium in the superconducting edge is large. The current flowing through the contact is a source of nonequilibrium quasiparticles, which are capable of multiplying by emitting phonons that are absorbed by Cooper pairs. The steady-state concentration \emph{n} of excitations in the superconducting vicinity of the contact is determined by the ratio of the rate of generation to the rate of recombination (escape) of the quasiparticles. The band gap $\Delta$ near the contact and the excess current, which is proportional to the band gap, depend on \emph{n}. When the critical injection power is reached, the state of the superconductor near the contact changes abruptly the excess current decreases by several per cent and the differential resistance grows. The position of the critical peculiarity in the point-contact spectrum on the \emph{V} axis is determined by the dependence of the quasiparticle recombination rate on the external parameters $(T, H)$. The shape of the spectrum $V>{{V}_{c}}$ is also determined by the dependence of the excitation relaxation rate on $eV$, $T$, and $H$. With increasing $eV$, the injection grows but at the same time so does the relaxation rate. As a result of the latter circumstance the spectra in the $S$ state are shifted down along the ordinate axis relative to the $N$ spectra. At not very high $eV$ an increase in the relaxation rate leads the injection and, therefore, in some range of $eV$ the band gap $\Delta$ has a tendency to increase, causing the I-V characteristic to have an anomalous curvature. At high values of $eV$ the increase in the relaxation rate slows down and $\Delta$ begins to decrease, which is responsible for the usual sign of the second derivative. The influence of the magnetic field and the temperature can be explained in similar fashion.

Impurities and increased contact size enhance the role of nonequilibrium phonons, whose reabsorption determines, to a considerable degree, the shape and position of lines in the point-contact spectra in the $S$ state at low temperatures. Effects of reabsorption of nonequilibrium phonons, therefore, can explain the observed spectral peculiarities for dirty contacts \cite{Yanson3}, when the energy diffusion length is comparable with or smaller than the diameter of the contact and the spectrum in the $N$ state spreads considerably. In this case, nonequilibrium phonons are generated in the direct proximity of the plane of the contact, where electrons that "remember" their energy at a large distance from the aperture are encountered. In the superconductor these are electrons of the condensate, since the electrochemical potential of Cooper pairs remains constant in space up to distances of the order of the inelastic mean free paths of electron-phonon relaxation. Phonons with a maximum energy $\hbar \omega =V$, therefore, are created near the boundary between a clean normal metal and a dirty superconductor, or between two dirty superconductors, in a point-contact and they then increase the concentration of nonequilibrium quasiparticles, decrease $\Delta $, and, as a result, reduce the excess current. Such, apparently, is the nature of the phonon peculiarities observed in $S-c-S$ contacts of $Nb$ (Ref.\cite{Yanson3}), $NbSe_{2}$ (Ref.\cite{Bobrov3}), and $Nb_{3}Sn$ (Ref.\cite{Yanson6}), as well as in $S-c-N$ contacts of $Tc-Ag$ (Ref.\cite{Zakharov}).

As long as effects of band-gap suppression and the transition of the superconductor to a spatially inhomogeneous state can be ignored, the I-V characteristics of clean contacts are described by the theory \cite{Khlus1,Khlus2}. Such a situation obtains in the case of $Sn-Cu$ contacts \cite{Yanson4} (between metals with fairly large ${{l}_{\varepsilon }}$ and $\xi $).

\section{NOTATION}
Here ${{p}_{F}}$ and ${{v}_{F}}$ are the electron Fermi momentum and velocity; $\Delta $ is the energy gap; ${{l}_{i}}$ and ${{l}_{\varepsilon }}$ are the momentum-relaxation and energy-relaxation mean free paths of electrons; ${{p}_{c}}$ and ${{V}_{c}}$ are the critical power and voltage; ${{J}_{c}}$ is the rate of quasiparticle generation per unit volume; ${{\lambda }_{Q}}$ is the diffusion length of excess quasiparticles; $Q^*$ is the density of excess quasiparticle charge; $d=2a$ is the point-contact diameter; and ${{\tau }_{Q}}(\varepsilon )$ is the charge-imbalance relaxation time for quasiparticles of energy $\varepsilon $.

\end{document}